\documentclass[11pt]{article}

\newcommand{\ShowComment}{true}

\usepackage{verbatim}
\usepackage{float}
\usepackage{amsmath}
\usepackage{amsthm}
\usepackage{amssymb}
\usepackage{thmtools}
\usepackage{graphicx}
\usepackage{mathtools}
\usepackage{enumitem}
\usepackage{array}
\usepackage{xspace}

\usepackage{tikz, relsize}
\usepackage{lmodern}
\usetikzlibrary{positioning, shapes, shadows, arrows}
\usetikzlibrary{decorations.pathreplacing}
	\tikzstyle{abstract}=[circle, draw=black, fill=white]
\tikzstyle{labelnode}=[circle, draw=white, fill=white]
\tikzstyle{line} = [draw, -latex']
\tikzset{fontscale/.style = {font=\relsize{#1}}}

\usepackage{geometry}
\floatstyle{ruled}
\newfloat{algorithm}{tbp}{loa}
\providecommand{\algorithmname}{Algorithm}
\floatname{algorithm}{\protect\algorithmname}

\usepackage{xcolor}
\usepackage{nameref}
\definecolor{ForestGreen}{rgb}{0.1333,0.5451,0.1333}
\definecolor{DarkRed}{rgb}{0.8,0,0}
\definecolor{Red}{rgb}{1,0,0}
\usepackage[linktocpage=true,
pagebackref=true,colorlinks,
linkcolor=DarkRed,citecolor=ForestGreen,
bookmarks,bookmarksopen,bookmarksnumbered]
{hyperref}

\usepackage{mathtools} %

\usepackage{cleveref}

\declaretheorem[numberwithin=section,refname={Theorem,Theorems},Refname={Theorem,Theorems},name={Theorem}]{thm}
\declaretheorem[numberlike=thm,refname={Theorem,Theorems},Refname={Theorem,Theorems},name={Theorem}]{theorem}

\declaretheorem[numberlike=thm,refname={Lemma,Lemmas},Refname={Lemma,Lemmas},name={Lemma}]{lem}
\declaretheorem[numberlike=thm,refname={Corollary,Corollaries},Refname={Corollary,Corollaries},name={Corollary}]{cor}

\declaretheorem[style=definition,numberlike=thm,refname={Definition,Definitions},Refname={Definition,Definitions},name={Definition}]{defn}

\AtBeginDocument{%
\let\Dref\ref
\let\ref\Cref
}

\usepackage{footnote}
\makesavenoteenv{tabular}
\makesavenoteenv{table}

\makeatletter
\newcommand\footnoteref[1]{\protected@xdef\@thefnmark{\labelcref{#1}}\@footnotemark}
\makeatother

\ifdefined\ShowComment
\newcommand{\todo}[1]{{\bf \color{red} TODO: #1}}

\newcommand{\danupon}[1]{{\color{red} DN: #1}}
\newcommand{\thatchaphol}[1]{\textcolor{blue}{TS: #1}}

\newcommand{\jl}[1]{{\bf \color{ForestGreen} JL: #1}}
\newcommand{\yg}[1]{{\bf \color{green} YG: #1}}

\newcommand{\yp}[1]{{\bf \color{green} RP: #1}}

\newcommand{\jnote}[1]{{\color{purple}{\sc\bf{[JC #1]}}}}
\newcommand{\takeout}[1]{{#1}}
\else
\newcommand{\todo}[1]{}
\newcommand{\danupon}[1]{}
\newcommand{\yp}[1]{}
\newcommand{\thatchaphol}[1]{}
\newcommand{\takeout}[1]{}

\newcommand{\yg}[1]{}

\newcommand{\jl}[1]{}
\newcommand{\jnote}[1]{}
\fi

\makeatother

\usepackage[english]{babel}

\newcommand{\myparskip}{3pt}
\parskip \myparskip
\newcommand{\set}[1]{\left\{ #1 \right\}}
\newcommand{\pset}{{\mathcal{P}}}

\newcommand{\hset}{{\mathcal{H}}}
\newcommand{\hG}{\hat G}

\global\long\def\cong{{\ensuremath{\eta}}}

\newcommand{\BCut}{{\sf BalCutPrune}\xspace}

\global\long\def\vol{\mathrm{Vol}}

\global\long\def\poly{\mathrm{poly}}
\global\long\def\polylog{\mathrm{polylog}}
\global\long\def\P{\mathcal{\mathcal{P}}}

\global\long\def\Otil{\tilde{O}}

\newcommand\ww{\boldsymbol{\mathit{w}}}

\newcommand\xx{\boldsymbol{\mathit{x}}}

\newcommand\eps{\epsilon}

\title{Deterministic Weighted Expander Decomposition\\in Almost-linear Time}

\author{
Jason Li\\
CMU\footnote{supported in part by NSF awards CCF-1907820, CCF-1955785, and CCF-2006953}
\and
Thatchaphol Saranurak\\
University of Michigan
}

\begin{document}

\makeatletter{\renewcommand*{\@makefnmark}{} \makeatother}

\maketitle

\abstract{
In this note, we study the expander decomposition problem in a more general setting where the input graph has positively weighted edges and nonnegative demands on its vertices. We show how to extend the techniques of \cite{ChuzhoyGLNPS20}
to this wider setting, obtaining a deterministic algorithm for the problem in almost-linear time. }

\pagenumbering{gobble}

\newpage

\pagenumbering{arabic}

\section{Introduction}

An \emph{$(\epsilon,\phi)$-expander decomposition} of a graph $G=(V,E)$
is a partition $\P=\{V_{1},\dots,V_{k}\}$ of the set $V$ of vertices, such that for all $1\leq i\leq k$, the conductance of graph $G[V_i]$ is at least $\phi$, and $\sum_{i-1}^k\delta_{G}(V_{i})\le\epsilon\vol(G)$. This decomposition was introduced in \cite{KannanVV04,GoldreichR99} and has been used as a key tool in many applications, including the ones mentioned in this paper.

Spielman and Teng \cite{SpielmanT04} provided the first near-linear time algorithm, whose running time is $\Otil(m/\poly(\epsilon))$, for computing a \emph{weak} variant of the $(\epsilon,\epsilon^2 / \poly(\log n))$-expander decomposition, where, instead of ensuring that each resulting graph $G[V_i]$ has high conductance, the guarantee is that for each such set $V_i$ there is some larger set $W_i$ of vertices, with $V_i\subseteq W_i$, such that $\Phi(G[W_i]) \ge \epsilon^2 / \poly(\log n)$.
 This caveat was first removed in \cite{NanongkaiS17}, who showed an algorithm for computing  an $(\epsilon,\epsilon / n^{o(1)})$-expander decomposition in time $O(m^{1+o(1)})$ (we note that \cite{Wulff-Nilsen17} provided similar results with somewhat weaker parameters). More recently, \cite{SaranurakW19} provided an algorithm for computing $(\epsilon,\epsilon / \poly(\log n))$-expander decomposition in  time $\Otil(m/\epsilon)$. Unfortunately, all algorithms mentioned above are randomized. 

Recently, a superset of the authors~\cite{ChuzhoyGLNPS20} obtained the first deterministic algorithm for computing an  $(\epsilon,\epsilon / n^{o(1)})$-expander decomposition in $m^{1+o(1)}$ time, which immediately implied near-optimal deterministic algorithms for many fundamental optimization problems, from dynamic connectivity to $(1+\epsilon)$-approximate max-flow in undirected graphs. While the expander decomposition algorithm only works for unweighted graphs, the applications can be adapted to work on weighted graphs via problem-specific reductions to the weighted case. However, since their initial work, further applications of expander decomposition have been discovered which require more sophisticated settings for the expander decomposition primitive itself, including weighted graphs~\cite{Li21} and even custom, arbitrary ``demands" on the vertices~\cite{LP20}.

In this note, we provide a fast, general-purpose expander decomposition algorithm that works for the widest setting known thus far: weighted graphs with custom demands on the vertices. While our algorithm is deterministic,  we remark that even a \emph{randomized} almost-linear-time algorithm in this setting was never explicitly shown before in the literature.

\subsection{Preliminaries from \cite{ChuzhoyGLNPS20}}

In this section, we introduce notation, definitions, and results from~\cite{ChuzhoyGLNPS20} relevant to this note.

All graphs considered in this paper are positively weighted and undirected.
Given a graph $G=(V,E,w)$, for every vertex $v\in V$, we denote by $\deg_G(v)$ the sum of weights of edges incident to $v$ in $G$. For any set $S\subseteq V$ of vertices of $G$,
the \emph{volume} of $S$ is the sum of degrees of all nodes in $S$: $\vol_{G}(S)=\sum_{v\in S}\deg_{G}(v)$. %
For an edge $e\in E$, we denote by $w(e)$ the weight of edge $w$, and for a subset $F\subseteq E$ of edges, we define $w(F)=\sum_{e\in F}w(e)$.

We use standard graph theoretic notation: for two subsets $A,B\subseteq V$ of vertices of $G$, we denote by $E_{G}(A,B)$ the set of all edges with one endpoint in $A$ and another in $B$. We sometimes write $w(A,B) = w(E_G(A,B))$. 
Assume now that we are given a subset $S$ of vertices of $G$. We denote by $G[S]$  the subgraph  of $G$ induced by $S$. We also denote $\overline S=V\setminus S$, and $G-S=G[\overline S]$.

An important subroutine in our algorithm is computing a \emph{spectral sparsifier} of a graph, defined below.
\begin{defn}[Spectral sparsifier]
The Laplacian
$L_{G}$ of $G$ is a matrix of size $n\times n$ whose entries are defined as follows: 
\[
L_{G}(u,v)=\begin{cases}
0 & u\neq v, (u,v)\not\in E\\
-\ww_{uv} & u\neq v, (u,v)\in E\\
\sum_{\stackrel{(u,u')\in E:} {u\neq u'}}\ww_{uu'} & u=v.
\end{cases}
\]
 We say that a graph $H$ is an \emph{$\alpha$-approximate spectral
sparsifier} for $G$ iff for all $\xx\in\mathbb{R}^{n}$, $\frac{1}{\alpha}\xx^{\top}L_{G}\xx\le \xx^{\top}L_{H}\xx\le\alpha\cdot \xx^{\top}L_{G}\xx$ holds. 
\end{defn}

\begin{thm}[Corollary~6.4 of \cite{ChuzhoyGLNPS20}]\label{cor:sparsifier}
There is a deterministic algorithm, that we call $\mathtt{SpectralSparsify}$ that,
given an undirected
$n$-node $m$-edge graph $G=(V,E,\ww)$ with edge weights in the range $[1,U]$, and a parameter $1\le r \le O(\log m)$, computes a $(\log m)^{O(r^2)}$-approximate
spectral sparsifier $H$ for $G$, with $|E(H)|\leq O\left(n \log n\log U\right)$, in time\linebreak
$O\left ( m^{1+O(1/r)}\cdot (\log m)^{O(r^2)}  \log U\right )$. %
\end{thm}

\newcommand{\f}{\frac}
\newcommand{\cd}{\cdot}
\newcommand{\bn}{\binom}
\newcommand{\sr}{\sqrt}
\newcommand{\cds}{\cdots}
\newcommand{\lds}{\ldots}
\newcommand{\vds}{\vdots}
\newcommand{\dds}{\ddots}
\newcommand{\pge}{\succeq}
\newcommand{\ple}{\preceq}
\newcommand{\sm}{\setminus}
\newcommand{\s}{\subseteq}
\newcommand{\su}{\supseteq}

\newcommand{\sumni}{\sum_{n=1}^\infty}
\newcommand{\sumin}{\sum_{i=1}^n}
\newcommand{\bigcupni}{\bigcup_{n=1}^\infty}
\newcommand{\bigcupin}{\bigcup_{i=1}^n}
\newcommand{\bigcapni}{\bigcap_{n=1}^\infty}
\newcommand{\bigcapin}{\bigcap_{i=1}^n}

\newcommand{\BE}{\begin{enumerate}}
\newcommand{\EE}{\end{enumerate}}
\newcommand{\im}{\item}
\newcommand{\BI}{\begin{itemize}}
\newcommand{\EI}{\end{itemize}}
\def\BAL#1\EAL{\begin{align*}#1\end{align*}}
\def\BALN#1\EALN{\begin{align}#1\end{align}}
\def\BG#1\EG{\begin{gather}#1\end{gather}}

\newcommand{\Sum}{\displaystyle\sum\limits}
\newcommand{\Prod}{\displaystyle\prod\limits}
\newcommand{\Int}{\displaystyle\int\limits}
\newcommand{\Lim}{\displaystyle\lim\limits}
\newcommand{\Max}{\displaystyle\max\limits}
\newcommand{\Min}{\displaystyle\min\limits}

\newcommand{\logn}{\log n}

\newcommand{\dx}{\frac d{dx}}
\newcommand{\dy}{\frac d{dy}}
\newcommand{\dz}{\frac d{dz}}
\newcommand{\dt}{\frac d{dt}}

\newcommand{\inv}{^{-1}}

\newcommand{\R}{\mathbb R}
\newcommand{\Z}{\mathbb Z}
\newcommand{\F}{\mathbb F}
\newcommand{\C}{\mathbb C}
\newcommand{\N}{\mathbb N}
\newcommand{\Q}{\mathbb Q}

\newcommand{\al}{\alpha}
\newcommand{\be}{\beta}
\newcommand{\el}{\ell}

\newcommand{\Ra}{\Rightarrow}

\newcommand{\lf}{\lfloor}
\newcommand{\rf}{\rfloor}
\newcommand{\lc}{\lceil}
\newcommand{\rc}{\rceil}

\newcommand{\lp}{\left(}
\newcommand{\rp}{\right)}
\newcommand{\lb}{\left[}
\newcommand{\rb}{\right]}
\newcommand{\lmt}{\left[\begin{matrix}}
\newcommand{\rmt}{\end{matrix}\right]}

\newcommand{\BT}{\begin{theorem}}
\newcommand{\ET}{\end{theorem}}
\newcommand{\BL}{\begin{lem}}
\newcommand{\EL}{\end{lem}}
\newcommand{\BD}{\begin{defn}}
\newcommand{\ED}{\end{defn}}

\newcommand{\BP}{\begin{proof}}
\newcommand{\EP}{\end{proof}}

\newcommand{\tO}{\tilde{O}}

\newcommand{\thml}[1]{\label{thm:#1}}
\newcommand{\thmm}[1]{\Cref{thm:#1}}
\newcommand{\leml}[1]{\label{lem:#1}}
\newcommand{\lemm}[1]{\Cref{lem:#1}}
\newcommand{\defnl}[1]{\label{def:#1}}
\newcommand{\para}{\paragraph}
\newcommand{\bd}{\mathbf d}
\newcommand{\WBCut}{\textsf{\textup{WeightedBalCutPrune}}}
\newcommand{\weighted}{\texttt{\textup{WeightedBalCut}}}
\newcommand{\rooted}{\texttt{\textup{RootedTreeBalCut}}}

\section{Weighted \BCut and  Expander Decomposition}

In our setting, 
every vertex $v\in V(G)$ has a non-negative demand $\bd(v)$ that is independent of the edge weights. 
As usual, 
the demand of a set $S\subseteq V(G)$ of vertices is $\bd(S)=\sum_{v\in S}\bd(v)$. 
Given a subset $S\subseteq V$ of vertices, we denote by  $\bd_{|S}$ the vector $\bd$ of demands restricted to the vertices of $S$.  We start by defining a weighted variant of sparsity and of expander decomposition.

\BD[Weighted Sparsity]
Given a graph $G=(V,E)$ with non-negative weights $w(e)\geq E$ on its edges $e\in E$, and non-negative demands $\bd(v)\geq 0$ on its vertices $v\in V$, the \emph{$\bd$-sparsity} of a subset $S\s V$ of vertices with $0<\bd(S)<\bd(V)$ is:
\[ \Psi_G^\bd(S)= \f{w(E_G(S,V\sm S))}{\min\{\bd(S),\bd(V\sm S)\}} .\]
The \emph{$\bd$-sparsity} of graph $G$ is  $\Psi^\bd(G)=\min_{S\subseteq V:0<\bd(S)<\bd(V)}\Psi_G^\bd(S)$.
\ED

Observe that if $w(e)=1$ for all $e\in E$ and $\bd(v)=\deg(v)$ for all $v\in V$, then this definition is exactly the conductance of the graph. Here, we use the term \emph{sparsity} instead of conductance because traditionally, sparsity concerns the number of vertices in the denominator of the ratio, while conductance uses volume which is closely related to the number of edges. However, for lack of an alternative term, we will stick with the term \emph{expander} to describe a graph of high weighted sparsity. We now define an expander decomposition for the weighted sparsity, which generalizes the standard definition for conductance.

\BD[Weighted Expander Decomposition]
Given a graph $G=(V,E,w,\bd)$ with non-negative weights $w(e)\geq 0$ on its edges $e\in E$, and non-negative demands $\bd(v)\geq 0$ on its vertices $v\in V$, a \emph{$(\epsilon,\psi)$-expander decomposition} of $G$ is a partition $\P=\{V_{1},\dots,V_{k}\}$ of the set $V$ of vertices, such that:
 \BE
 \im For all $1\leq i\leq k$, the graph $G[V_i]$ has $\bd|_{V_i}$-sparsity at least $\psi$, and %
 \im $\sum_{i-1}^k w ( E_G(V_{i}, V\sm V_i) ) \le\epsilon \bd(V)$.
 \EE
\ED

Similarly to~\cite{ChuzhoyGLNPS20}, the key subroutine of our expander decomposition algorithm is solving the following $\WBCut$ problem, a generalization of $\BCut$ from~\cite{ChuzhoyGLNPS20} that allows both weighted edges and ``demands" on the vertices.

\begin{defn}[$\WBCut$ problem]
        The input to the $\alpha$-approximate $\WBCut$ problem is a graph $G=(V,E)$ with non-negative weights $w(e)\geq 0$ on edges $e\in E$, a nonzero vector $\bd\in\R^V_{\ge0}$ of demands, a sparsity parameter $0<\psi\leq 1$, and an approximation factor $\alpha$. The goal is to compute a partition $(A,B)$ of $V(G)$ (where possibly $B=\emptyset$),
        with $w(E(A,B))\leq \alpha \psi\cdot \min\{\bd(A),\bd(B)\}$,\footnote{We remark that this guarantee is stronger than what we would obtain if we directly translated $\BCut$ from~\cite{ChuzhoyGLNPS20}. The latter only requires that $|E(A,B)|\le\al\psi\cd\vol(G)$ in their setting, which would translate to $w(E(A,B))\le\al\psi\cd\bd(V)$ in our setting.} such that one of the following hold:
        either
        \begin{enumerate}[noitemsep]
                \item \textbf{(Cut)} $\bd(A),\bd(B)\ge \bd(V)/3$; or
                \item \textbf{(Prune)} $\bd(A)\geq \bd(V)/2$, and $\Psi^{\bd|_A}(G[A])\ge\psi$.%
        \end{enumerate}
\end{defn}

The main technical result of this note is the following algorithm for $\WBCut$.%

\BT\thml{WBCut}
        There is a deterministic algorithm that, given an $m$-edge connected graph $G=(V,E)$ with edge weights $1\leq w(e)\leq U$ for all $e\in E$ and demands $\bd(v)\in\{0\}\cup [1,U]$ for all $v\in V$ that are not all zero,
        together with parameters $0<\psi\leq 1$ and $r\geq 1$, solves the $(\log^{O(r^4)}m)$-approximate $\WBCut$ problem in time $m\cd(mU)^{O(1/r)}$.
\ET

We provide the proof of \thmm{WBCut} in the following subsections. Before we do so, we obtain the following corollary, whose proof follows similarly to the reduction from expander decomposition to $\BCut$ in~\cite{ChuzhoyGLNPS20}. %
For completeness, we include the proof in \Cref{sec:completing}. %

\begin{cor} \label{cor:exp-decomp-sec8}
There is a deterministic algorithm that, given an $m$-edge graph
$G=(V,E)$ with weights $1\leq w(e)\leq U$ on its edges $e\in E$, and demands $\bd(v)\in\{0\}\cup[1,U]$ for its vertices $v\in V$ that are not all zero,
        together with a parameter $\epsilon\in(0,1]$ and $r\ge 1$, computes
a $\left(\epsilon,\psi\right )$-expander decomposition
of $G$, for $\psi=\epsilon/ \big( \log^{O(r^4)}m \log U \big) $, in time $m\cd(mU)^{O(1/r)}\log(mU)$.
\end{cor}

\subsection{Weighted Most-Balanced Sparse Cut}

We first define the Weighted Most-Balanced Cut problem, and provide a bi-criteria approximation algorithm for it, this time based on recursively applying the $j$-tree framework of Madry~\cite{Madry10}. In \Cref{sec:completing}, we then show our algorithm for Weighted Most-Balanced Cut can be used in order to approximately solve the $\WBCut$ problem.

\BD[$(s,b)$-most-balanced $\psi$-sparse cut]\label{def:most-balanced-s}
Given a graph $G=(V,E)$ and parameters $s,b\ge1$, a set $S\s V$ with $\bd(S)\le\bd(V)/2$ is a \emph{$(s,b)$-most-balanced $\psi$-sparse cut} if it satisfies:
 \BE
 \im $w(S,V\sm S)\le\psi\cd\bd(S)$.
 \im Define $\psi^*:=\psi/s$ and let $S^*\s V$ be the set with maximum $\bd(S^*)$ out of all sets $S'$ satisfying $w(S',V\sm S')\le\psi^*\cd\min\{\bd(S'),\bd(V\sm S')\}$ and $\bd(S')\le\bd(V)/2$. Then, $\bd(S)\ge\bd(S^*)/b$.
 \EE
\ED

Let us first motivate why we consider a completely different recursive framework based on recursive $j$-trees~\cite{Madry10} instead of the recursive KKOV cut-matching game framework \cite{KhandekarKOV2007cut} as used in  \cite{ChuzhoyGLNPS20}. This is because KKOV recursion scheme does not
generalize easily to the weighted setting.
The main issue that in a weighted graph, the flows constructed by the matching player cannot be decomposed into a small number of paths; the only bound we can prove is at most $m$ paths by standard flow decomposition arguments. Hence, the graphs constructed by the cut player are not any sparser, preventing us from obtaining an efficient recursive bound.
Madry's $j$-tree framework,
on the other hand, generalizes smoothly to weighted instances and
can even be adapted to solve the sparsest cut problem with general demands, for which
Madry provided efficient randomized algorithms in his original paper~\cite{Madry10}.

Below, we give a high-level description of Madry's approach. But first, let us state the definition of $j$-trees as follows.
\begin{defn} \label{def:jtree}
        A graph $G$ is a $j$-tree if it is a union of:
        \begin{itemize}
                \item a subgraph $H$ of $G$ (called the \emph{core}), induced by a set $V_{H}$ of at most $j$ vertices; and
                \item a forest (that we refer to as \emph{peripheral forest}), where each connected component of the forest contains exactly one vertex of $V_{H}$.
                For each core vertex $v \in V_{H}$, we let $T_G(v)$ denote the unique tree in the peripheral forest that contains $v$. When the $j$-tree $G$ is
                unambiguous, we may use $T(v)$ instead.%
        \end{itemize}
\end{defn}

In Madry's approach, the input graph is first decomposed into a small number of
$j$-trees (formally stated in \Cref{thm:madrythm}), so that it suffices to solve the problem on each $j$-tree and take the best
solution.
For a given $j$-tree, one key property of the generalized sparsest cut
problem is that either the optimal solution only cuts edges of the core, or it only cuts edges of the peripheral forest.
Therefore, the algorithm can solve two separate problems,
one on the core and one on the peripheral forest.
The former becomes a recursive call on a graph of $j$ vertices,
and the latter simply reduces to solving the problem on a tree.

This same strategy almost directly translates over to the Weighted Most-Balanced Cut problem. The main additional difficulty is in ensuring the additional \emph{balanced} guarantee in our Weighted Most-Balanced Cut problem, which is the biggest technical component of this section. We remark that our algorithm for computing the weighted most-balanced sparse cut is a modification of the algorithm in Section~8~of~\cite{GaoLNPSY19}. In particular, the algorithms $\weighted$ and $\rooted$ presented below are direct modifications of Algorithm~4 and Algorithm~5 in Section~8~of~\cite{GaoLNPSY19}, respectively. Still, we assume no familiarity with that paper and make no references to it.

We now state formal definition of graph embedding and Madry's decomposition theorem for $j$-trees below.

\begin{defn}\label{def:embed}
        Let $G$, $H$ be two graphs with $V(G)= V(H)$. An \emph{embedding} of $H$ into $G$ is a collection $\pset=\set{P(e)\mid e\in E(H)}$ of paths in $G$, such that for each edge $e\in E(H)$, path $P(e)$ connects the endpoints of $e$ in $G$. We say that the embedding causes \emph{congestion} $\cong$ iff every edge $e'\in E(G)$ participates in at most $\cong$ paths in $\pset$. %
\end{defn}

\begin{lem}[\cite{Madry10-jtree}] \label{thm:madrythm}
        There is a deterministic algorithm that, given an edge-weighted graph $G=(V, E, \ww)$ with $|E|=m$ and capacity ratio $U=\frac{\max_{e\in E}\ww_e}{\max_{e\in E}\ww_e}$, together with a parameter $t \geq 1$, computes, in time $\tilde{O}(tm)$, a distribution $\set{\lambda_i}_{i=1}^{t}$ over a collection of $t$ edge-weighted graphs $G_1,\ldots,G_{t}$, where for each $1\leq i\leq t$, $G_i=(V, E_i, \ww_i)$, and the following hold:
        \begin{itemize}
                \item for all $1\leq i\leq t$, graph $G_i$ is an $(\frac{m\log^{O(1)}m\log U}t)$-tree, whose core contains at most $m$ edges;
                \item for all $1\leq i\leq t$, $G$ embeds  into $G_i$ with congestion $1$; and %
                \item the graph that's the average of these graphs over the distribution,
                $\tilde G= \sum_{i} \lambda_i G_i$ can be embedded into $G$ with congestion $O(\log m (\log\log m)^{O(1)})$.
        \end{itemize}
                Moreover, the capacity ratio of each $G_i$ is at most $O(mU)$.
\end{lem}

In particular, \Cref{def:embed,thm:madrythm} imply that, for any cut $(S,V\setminus S)$, we have that
$w(E_{G_i}(S,V \setminus S))\ge  w(E_G(S,V \setminus S))$ for all $i$, and there exists $i$ where $w(E_{G_i}(S,V \setminus S))\le \be \cd w(E_G(S,V \setminus S))$. This is the fact that we will use later.

Our algorithm $\weighted$  first invokes \Cref{thm:madrythm} to approximately decompose the input graph $G$ into $t$ many $j$-trees, where $j=O(m/t)$ and $t$ is small (say, $m^\eps$ for some constant $\eps>0$). Since the distribution of $j$-trees approximates $G$, it suffices to solve the Weighted Most-Balanced Cut problem on each $j$-tree separately and take the best overall. For a given $j$-tree $H$, the algorithm computes two types of cuts---one that only cuts edges in the core of $H$, and one that only cuts edges of the peripheral forest of $H$---and takes the one with better weighted sparsity. In our analysis (specifically \Cref{lem:either}), we prove our correctness by showing that for any cut $S$ of the $j$-tree $H=(V_H,E_H)$, there exists a cut $S'$ that
\begin{enumerate}
        \item either only cuts core edges or only cuts peripheral edges, and
        \item has weighted sparsity and balance comparable to those of $H$, i.e.,
        $w_H(E_H(S',V_H\sm S')) \le O(w_H(E_H(S,V_H\sm S)))$
        and
        $\bd(S') \ge \Omega(\bd(S))$.
\end{enumerate}

To compute the best way to cut the core, the algorithm first contracts all edges in
the peripheral forest, summing up the demands on the contracted vertices. This leaves
a graph of $j=O(m/t)$ vertices, but the number of edges can still be $\Omega(m)$. To
ensure the number of edges also drops by a large enough factor, the algorithm
sparsifies the core using \Cref{cor:sparsifier},
computing a sparse graph with only $\tO(m/t)$ edges that $\al$-approximates
all cuts of the core for some $\al=(\log{m})^{O(r^2)}$.
Finally, the algorithm recursively solves the problem on
the sparsified core. The approximation factor blows up by $\polylog(m)$ per recursion
level, but the number of edges decreases by roughly $t=m^{\eps}$, so over the
$O(1/\eps)$ recursion levels, the overall approximation factor becomes $(\log m)^{O(r^2/\epsilon)}$, which is $n^{o(1)}$ appropriate choices of $r$ and $\epsilon$.

The algorithm for cutting the peripheral forest is much simpler and non-recursive. The algorithm first contracts the core of $H$, obtaining a tree in which to compute an approximate Weighted Most-Balanced Cut. Then, $\rooted$ roots the tree at an appropriately chosen ``centroid" vertex and greedily adds subtrees of small enough sparsity into a set $S$ until either $\bd(S)$ is large enough, or no more sparse cuts exist.

\begin{algorithm}[H]
 $\weighted ( G, \psi, \psi^*, b )$ with $\psi \ge \psi^*$ and $b\ge1$, and $G$ has demands $\bd$:
\BE

\im Fix an integer $r\ge1$ and parameter $t = \left\lceil m_0^{1/r} (\log m)^{O(1)} \log^2 U \right\rceil$, where $m_0$ is the number of edges in the \emph{original} input graph to the recursive algorithm, $m\le m_0$ is the number of edges of the input graph $G$ to the current recursive call, and $U$ is the capacity ratio of  $G$.

\im \label{line:def beta} Fix parameters  $\al =(\log m)^{O(r^2)}$ as the approximation factor from \Cref{cor:sparsifier}, and $\be =O(\log m(\log\log m)^{O(1)})$ as the congestion factor from \Cref{thm:madrythm}.

\im Compute $O(m/t)$-trees $G_1,\lds,G_t$ using \Cref{thm:madrythm} with $G$ and $t$ as input. For each $i$, let $K_i$ denote the vertex set in the core of $G_i$

\im For each $i\in[t]$:
 \BE

\im $H_i \gets G_i[K_i]$ with demands $\bd_{H_i}$ on $K_i$ as $\bd_{H_i}(v)=\sum_{u\in V(T_{G_i}(v))}\bd(u)$ (so that $\bd_{H_i}(K_i)=\bd(V)$).

\im $H_i'\gets \al $-approximate spectral sparsifier of $H_i$ (with the same demands)

\im $S'_{H_i}\gets \weighted(H_i',\, \psi/\al ,\, 3\al \be \psi^*,\, b/3)$ \label{line:a1}

\im $S_{H_i} \gets S'_{H_i}$ with each vertex $v$ replaced with $V(T_{G_i}(v))$ (see \Cref{def:jtree})\label{line:uncontract H}

\im Construct a tree $T_i=(V_{T_i},E_{T_i},w_{T_i})$ with demands $\bd_{T_i}$ as follows: Starting with $G_i$, contract $K_i$ into a single vertex $k_i$ with demand $\bd(K_i)$. All other vertices have demand $\bd(v)$ (so that $\bd_{T_i}(V_{T_i})=\bd(V)$).\label{line:a2}

\im Root $T_i$ at a vertex $r_i\in V_{T_i}$ such that every subtree rooted at a child of $r_i$ has total weight at most $\bd_{T_i}(V)/2=\bd(V)/2$. \label{line:a3}

\im $S'_{T_i}\gets \rooted(T_i,r_i,\psi)$

\im $S_{T_i}\gets S'_{T_i}$ with the vertex $k_i$ replaced with $K_i$ if $k_i\in S'_{T_i}$\label{line:uncontract T}
\EE

\im Of all the cuts $S=S_{H_i}$ or $S=S_{T_i}$ computed satisfying $w(S,V\sm S)\le\psi\cd\min\{\bd(S),\bd(V\sm S)\}$, consider the set $S$ with maximum $\min\{\bd(S),\bd(V\sm S)\}$, and output $S$ if $\bd(S)\le\bd(V\sm S)$ and $V\sm S$ otherwise. If no cut $S$ satisfies $w(S,V\sm S)\le\psi\cd\min\{\bd(S),\bd(V\sm S)\}$, then return $\emptyset$.

\EE
\end{algorithm}

\begin{algorithm}[H]
$\rooted(T=(V_T,E_T,w_T),r,\psi_T)$:

\BE
\im[0.] Assumption: $T$ is a weighted tree with demands $\bd_T$. The tree is rooted at a root $r$ such that every subtree $V_u$ rooted at a vertex $u\in V_T\sm\{r\}$ has total demand $\bd_T(V_u)\le \bd_T(V_T)/2$.
\\Output: a set $S\s V_T$ satisfying the conditions of \Cref{lem:tree-alg}.

\im Find all vertices $u\in V_T\sm\{r\}$ such that if $V_u$ is the vertices in the subtree rooted at $u$, then $w_T(E[V_u,V_T\sm V_u])/\bd_T(V_u)\le2\psi_T$. Let this set be $X$.
\im Let $X^\uparrow$ denote all vertices $u\in X$ without an ancestor in $X$ (that is, there is no $v\in X\sm\{u\}$ with $u\in T_v$).
\im Starting with $S=\emptyset$, iteratively add the vertices $V_u$ for $u\in X^\uparrow$. If $\bd_T(S)\ge \bd_T(V_T)/4$ at any point, then terminate immediately and output $S$. Otherwise, output $S$ at the end.\label{line:t3}

\EE
\end{algorithm}

We now analyze our algorithm $\weighted$ by showing the following:

\BL\leml{analysis}
Fix parameters $b\ge6$, $\psi^*>0$, and $\psi\ge12\be\cd\psi^*$ for $\be$ as defined in Line \Dref{line:def beta} of $\weighted$ algorithm.
$\weighted$ outputs a $(\psi/\psi^*,b)$-most-balanced $(\psi,\bd)$-sparse cut.%
\EL

We now state our structural statement on cuts in $j$-trees: for each $j$-tree $G_i$, either the core $H_i$ contains a good balanced cut or the ``peripheral" tree $T_i$ (produced by contracting the core) does.

\BL\label{lem:either}
Fix $i\in[t]$, and let $S^*\s V$ be any cut with $\bd(S^*)\le \bd(V)/2$. For simplicity, define $K=K_i$, $k=k_i$, $T=T_i$, $r=r_i$, and $H=H_i$. One of the following must hold:
\BE
\im There exists a cut $S^*_T\s V_T$ in $T$ satisfying $w_T(E_T(S^*_T,V_T\sm S^*_T))\le w(E_{G_i}(S^*,V\sm S^*))$ and $\bd(S^*)/2\le \bd_T( S^*_T)\le 2\bd(V)/3$, and $S^*_T$ is the disjoint union of subtrees of $T$ rooted at $r$.
\im There exists a cut $S^*_H\s K$ in core $H$ satisfying $w_H(E_{H}(S^*_H,K\sm S^*_H))\le w(E_{G_i}(S^*,V\sm S^*))$ and  $\min \{ \bd_H(S^*_H) , \bd_H(K\sm S^*_H)\} \ge\bd(S^*)/3$.
\EE
\EL

The statement itself should not be surprising. If $S^*$ only cuts edges in the peripheral forest of $G_i$, then the cut survives when we contract the core $H$ to form the tree $T$, and its $\bd_T$-sparsity is the same as its original $\bd$-sparsity. Likewise, if $S^*$ only cuts edges in the core $H$, then the cut survives when we contract the all edges in the peripheral forest to form $K$, and its $\bd_H$-sparsity is the same as its original $\bd$-sparsity. The difficulty is handling the possibility that $S^*$ cuts both peripheral forest edges and core edges, which we resolve through some casework below.

\BP
We need a new notation. For a $j$-tree $G_i$ and a vertex $v$ on peripheral forest $F$, we define $c_{G_i}(v)$ as the unique vertex shared by $F$ and the core $H$ of $G_i$.%

\begin{figure}
\centering \includegraphics[scale=.7]{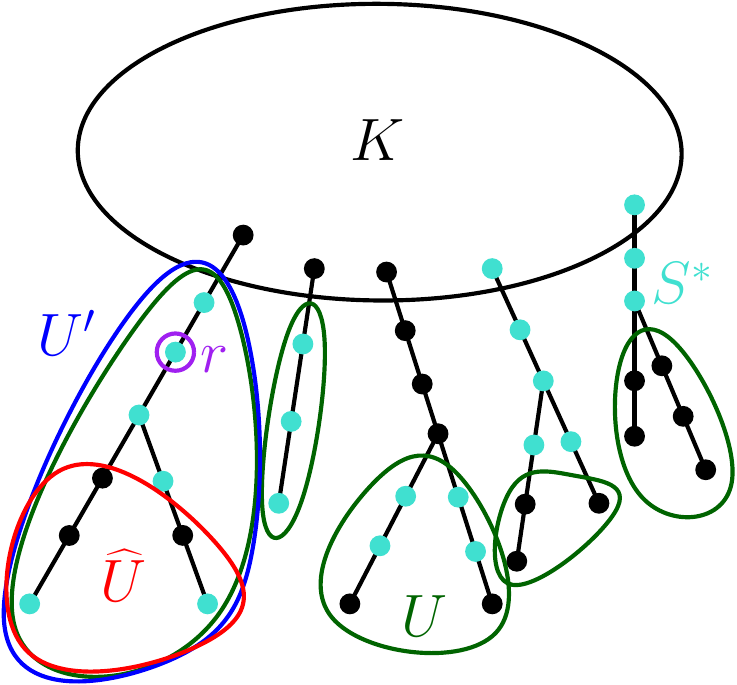} \qquad \includegraphics[scale=.7]{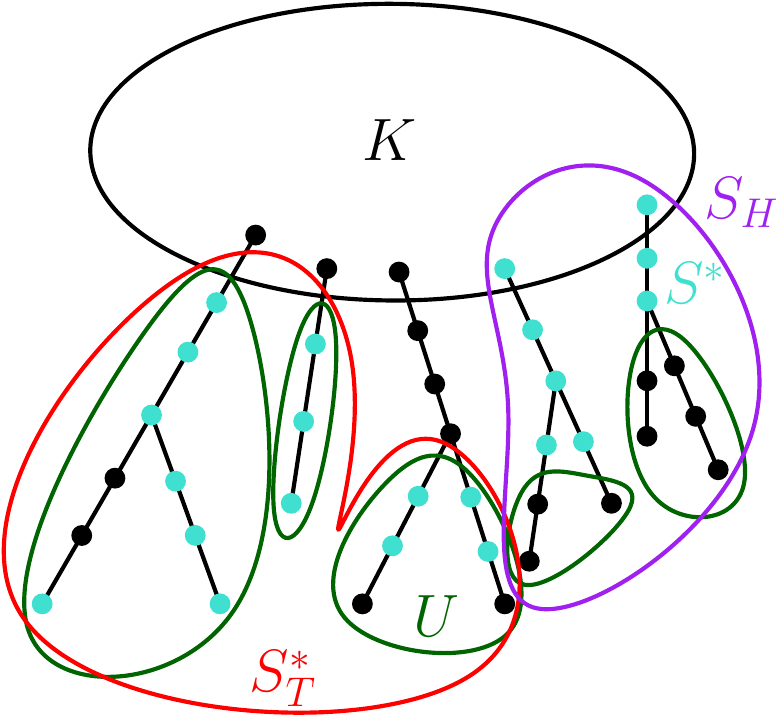}
\caption{
Left: Cases~1a~and~1b of \Cref{lem:either}. The set $S^*$ is the cyan vertices. Right: Cases~2a and 2b.
}
\label{fig:j}
\end{figure}

Let $S^*\s V$ the set as described in \ref{def:most-balanced-s} ($w(S^*,V\sm S^*)\le\psi^* \cd\min\{\bd(S^*),\bd(V\sm S^*)\}$). Let $U$ be the vertices $u\in V$ whose (unique) path to $c_{G_i}(u)$ in $F$ contains at least one edge in $E_{G_i}(S^*,V\sm S^*)$. In  \ref{fig:j}, $U$ is the set of vertices with green circle around. Note that $U\cap K=\emptyset$ and $E_{G_i}(U,V\sm U)\s E_{G_i}(S^*,V\sm S^*)$. 
Observe further that $U$ is a union of subtrees of $T$ rooted at $k$ (not $r$). This is because, when we root the tree $T$ at $k$, for each vertex $u\in U$, its entire subtree is contained in $U$.

\para{Case 1: $r\in U$.}
In this case, we will construct a cut in the tree $T$ to fulfill condition~(1).
        Let $F$ be the peripheral forest of $G_i$ (see \Cref{def:jtree}) and let $T'$ be the tree in $F$ that contains $r$. Define $U' = T' \cap U$ (\ref{fig:j} left). In words, $U'$ contains all vertices of $U$ in the tree of $F$ that contains $r$.
    Let us re-root $T$ at vertex $r$, so that the vertices in $V_T\sm U'$ now form a subtree. We now consider a few sub-cases based on  the size of $U'$.

\para{Case 1a: $r\in U$ and $\bd_T(U')\le3\bd(V)/4$.}Define $S^*_T\s V_T$ as $S^*_T:=V_T\sm U'$ . By our selection of $r$,
\[ \bd_T(S^*_T) =\bd_T(V_T\sm U')\le \f{\bd(V)}2 .\]
Moreover,
\[ \bd_T(S^*_T)=\bd_T(V_T\sm U')\ge \f{\bd(V)}4\ge \f{\bd(S^*)}2 \] and \[ E_T(S^*_T,V\sm S^*_T)\s E_{G_i}(U,V\sm U)\s E_{G_i}(S^*,V\sm S^*) ,\]
fulfilling condition~(1).%

\para{Case 1b: $r\in U$ and $\bd_T(U')\ge3\bd(V)/4$.}
Define $\widehat U$ as all vertices $u\in U'$ whose (unique) tree path to root
($r)$ contains at least one vertex not in $S^*$ (possibly $u$ itself).
As this set contains all vertices in $U'$ not in $S^{*}$,
we have $\widehat U\supseteq U'\sm S^*$, and in turn
\[
\bd_T\left(\widehat U\right)
\ge
\bd_T\left(U'\sm S^*\right)
=
\bd_T\left(U'\right) - \bd_T\left(U'\cap S^*\right)
\ge
\bd_T\left(U'\right)-\bd\left(S^*\right)
\ge
\frac{3\bd\left(V\right)}{4}-\frac{\bd\left(V\right)}{2}
=\frac{\bd(V)}{4}.
\]
Moreover, $\widehat U$ is a union
of subtrees of $T$ rooted at $r$ and satisfies
\[
E_{G_i}\left(\widehat U,V\sm\widehat U\right)
\s
E_{G_i}\left(S^*,V\sm S^*\right).
\]

By our choice of $r$, each subtree $T'$ of $U$ satisfies
$\bd_T(V(T'))\le\bd(V)/2$.
We perform one further case work based on the largest size of
one of these subtrees to show that we can find a tree cut
that satisfies condition~(1).
\begin{itemize}
\item If there exists a subtree $T'\s\widehat U$ with
$\bd_T(V(T'))\ge\bd(V)/4$, then set $S^*_T:=V(T')$.
\item
Otherwise, since $\bd_T(\widehat U)\ge\bd(V)/4$, we can greedily select a subset of
subtrees of $\widehat U$ with total $\bd(\cd)$ value in the range $[\bd(V)/4,\bd(V)/2]$,
and set $S^*_T$ as those vertices.
\end{itemize}
In both cases we have
\[
\frac{\bd\left(S^*\right)}{2}
\le
\frac{\bd(V)}{4}
\le
\bd_T\left(S^*_T\right)
\le
\frac{\bd\left(V\right)}{2}
\]
which gives the volume condition on $S^{*}$,
and the cut size bound follows from $w(E_T(S^*_T,V\sm S^*_T))\le w(
E_{G_i}(\widehat U,V\sm \widehat U))$.

\para{Case 2: $r\notin U$.} In this case, we will cut either the tree $T$ or the core $H$ depending on a few further sub-cases.

\para{Case 2a: $r\notin U$ and $\bd_T(U)\ge \bd(V)/6$.}
Since $r\notin U$, every subtree in $U$ has weight at most $\bd(V)/2$. Let $U'$ be a subset of these subtrees of total $\bd(\cd)$ value in the range $[\bd(V)/6,2\bd(V)/3]$.
Define the tree cut $S^*_T:=U'$, which satisfies 
\[ \f{\bd(V)}2\ge\bd_T(S^*_T)\ge \f{\bd(V)}6\ge\f{\bd(S^*)}3 \] and \[ E_T(S^*_T,V\sm S^*_T)\s  E_{G_i}(U,V\sm U)\s E_{G_i}(S^*,V\sm S^*),\] fulfilling condition~(1).

\para{Case 2b: $r\notin U$ and $\bd_T(U)<\bd(V)/6$.}
In this case, let $S:=S^*\cup U$, which satisfies \[ \bd(S^*)\le\bd(S)\le\bd(S^*)+\bd_T(U)\le\bd(S^*)+\bd(V)/6\le2\bd(V)/3 \] and $E_{G_i}(S,V\sm S)\s E_{G_i}(S^*,V\sm S^*)$. Next, partition $S$ into $S_H$ and $S_T^*$ according to \ref{fig:j}, where $S_H$ consists of the vertices of all connected components of $G_i[S]$ that intersect $K$, and $S_T^*:=S\sm S_H$ is the rest. We have 
\[ E_{G_i}(S_H,V\sm S_H)\s E_{G_i}(S^*,V\sm S^*) \qquad\text{and}\qquad E_{G_i}(S_T^*,V\sm S_T^*)\s E_{G_i}(S^*,V\sm S^*) .\]

Observe that $S^*_T$ is a tree cut, and $S_H$ is a core cut since it does not cut any edges of the peripheral forest. We will select either $S^*_T$ or $S_H$ based on one further case work.

Since $\bd(S^*_T)+\bd(S_H)=\bd(S)$, we can
case on whether $\bd_T(S_T^*)\ge\bd(S)/2$ or $\bd(S_H)\ge\bd(S)/2$. 
\BI
\im If $\bd_T(S_T^*)\ge\bd(S)/2$, then the set $S^*_T$ satisfies condition~(1).
 \im Otherwise, $\bd(S_H)\ge\bd(S)/2$.
Since $E_{G_i}(S_{H},V\sm S_{H})$ does not contain any edges in the peripheral forest $F$, we can ``contract" the peripheral forest to  obtain the set $S^*_H:=\{c_{G_i}(v):v\in S_H\}\s K$ such that $S_H$ is the vertices in the trees in $F$ intersecting $S^*_H$. This also means that $V\sm S_H$ is the vertices in the trees of $F$ intersecting $K\sm S^*_H$. It remains to show that $S^*_H$ fulfills condition~(2). We have 
\[ w_H(E_{H}(S^*_H,K\sm S^*_H))=w(E_{G_i}(S_H,V\sm S_H))\le w(E_{G_i}(S^*,V\sm S^*)) \] and
\[ \min \{ \bd_H(S^*_H) , \bd_H(K\sm S^*_H)\} = \min \{ \bd(S_H),\bd(V\sm S_H)\} .\]
It remains to show that $\min\{\bd(S_H),\bd(V\sm S_H)\}\ge\bd(S^*)/3$.
This is true because
$\bd(S_H)\ge\bd(S)/2\ge \bd(S^*)/2$ and
$\bd(S_H)\le\bd(S)\le2\bd(V)/3$ which means that $\bd(V\sm S_H) \ge \bd(V)/3 \ge \bd(S^*)/3$.
\EI
\EP

If the graph $H$ contains a good balanced cut, then intuitively, the demands $\bd_H$ are set up so that the recursive call on $H'$ will find a good cut as well. The lemma below shows that if the tree $T$ contains a good balanced cut, then $\rooted$ will perform similarly well.

\BL\label{lem:tree-alg}
$\rooted(T=(V_T,E_T,w_T),\bd_T,r,\psi_T)$ can be implemented to run in $O(|V_T|)$ time.
The set $S$ output satisfies $\psi_{T}^{\bd_T}(S)=w_T(E_T(S,V_T\sm S))/\min\{\bd_T(S),\bd_T(V_T\sm S)\}\le6\psi_T$.
Moreover, for any set $S^*$ with $w_T(E_T(S^*,V_T\sm S^*))/ \bd_T(S^*)\le \psi_T$ and  $ \bd_T(S^*)\le2 \bd_T(V_T)/3$, and which is composed of vertex-disjoint subtrees rooted at vertices in $T$, we have $\min\{ \bd_T(S), \bd_T(V_T\sm S)\}\ge  \bd_T(S^*)/3$.
\EL
\BP
Clearly, every line in the algorithm can be implemented in linear time, so the running time follows. We focus on the other properties.

Every set of vertices $V_u$ added to $S$ satisfies $w_T(E_T(V_u,V_T\sm V_u))/ \bd_T(V_u)\le2\psi_T$. Also, the added sets $V_u$ are vertex-disjoint, so $w_T(E_T(S,V_T\sm S)) = \sum_{V_u\s S}w_T(E_T(V_u,V_T\sm V_u))$. This means that $\rooted$ outputs $S$ satisfying $w_T(E_T(S,V_T\sm S))/ \bd_T(S)\le2\psi_T$. Since every set $V_u$ has total weight at most $ \bd_T(V_T)/2$, and since the algorithm terminates early if $ \bd_T(S)\ge  \bd_T(V_T)/4$, we have $ \bd_T(S)\le 3 \bd_T(V_T)/4$. This means that $\min\{ \bd_T(S), \bd_T(V_T\sm S)\}\ge \bd_T(S)/3$, so $w_T(E_T(S,V_T\sm S))/\min\{ \bd_T(S), \bd_T(V_T\sm S)\} \le 3w_T(E_T(S,V_T\sm S))/ \bd_T(S)\le6\psi_T$.

It remains to prove that $S$ is balanced compared to $S^*$. There are two cases. First, suppose that the algorithm terminates early. Then, as argued above, $\min\{ \bd_T(S), \bd_T(V_T\sm S)\}\ge  \bd_T(V_T)/4$, which is at least $(2 \bd_T(V_T)/3)/3\ge  \bd_T(S^*)/3$, so  $\min\{ \bd_T(S), \bd_T(V_T\sm S)\}\ge  \bd_T(S^*)/3$.

Next, suppose that $S$ does not terminate early. From the assumption of $S^*$, there are sets $S^*_1,\lds,S^*_\el$ of vertices in the (vertex-disjoint) subtrees that together compose $S^*$, that is, $\bigcup_iS^*_i=S^*$. Note that $E_T(S^*_i,V_T\sm S^*_i)$ is a single edge in $E_T$ for each $i$. Suppose we reorder the sets $S^*_i$ so that $S^*_1,\lds, S^*_q$ are the sets that satisfy $w_T(E_T(S^*_i,V_T\sm S^*_i))/ \bd_T(S^*_i)\le2\psi_T$. From the assumption on $S^*$, we have $w_T(E_T(S^*,V_T\sm S^*))/ \bd_T(S^*)\le \psi_T$, by a Markov's inequality-like argument, we must have $\sum_{i\in[q]} \bd_T(S^*_i) \ge (1/2)\sum_{i\in[\el]} \bd_T(S^*_i)= \bd_T(S^*)/2$. Observe that by construction of $X^\uparrow$, each of the subsets $S^*_1,\lds, S^*_q$ is inside $V_u$ for some $u\in X^\uparrow$. Therefore, the set $S$ that $\rooted$ outputs satisfies $ \bd_T(S)\ge\sum_{i\in[q]} \bd_T(S^*_i)\ge  \bd_T(S^*)/2$. 
\EP

Finally, we prove \lemm{analysis}:

\BP[Proof (\lemm{analysis})]
Let $S^*\s V$ be the set for $G$ as described in \Cref{def:most-balanced-s} with parameters $s=\psi/\psi^*$ and $b$; that is, it is the set with maximum $\bd(S^*)$ out of all sets $S'$ satisfying  $\Psi_{G}^{\bd}(S')\le\psi ^*$ and $\bd(S')\le\bd(V)/2$. If $\bd(S^*)=0$, then the output of $\weighted$ always satisfies the definition of $(s,b)$-most-balanced $\psi$-sparse cut, even if it outputs $\emptyset$. So for the rest of the proof, assume that $\bd(S^*)>0$, so that $\Psi_G^\bd(S^*)$ and $\Psi_{G_i}^\bd(S^*)$ are well-defined.

By \Cref{thm:madrythm}, there exists $i\in[t]$ such that $w(E_{G_i}(S^*,V \setminus S^*))\le \be \cd w(E_G(S^*,V \setminus S^*))$, which means that
\[ \Psi_{G_i}^\bd(S^*)\le\be \cd \Psi_G^\bd(S^*) \le \be \cd\psi^* .\]
For the rest of the proof, we focus on this $i$, and define $K=K_i$, $H=H_i$, and $T=T_i$. We break into two cases, depending on which condition of \lemm{either} is true:
 \BE
 \im Suppose condition~(1) is true for the cut $S^*_T$. Then, since $w_T(E_T(S^*_T,V_T\sm S^*_T))\le w(E_{G_i}(S^*,V\sm S^*))$ and $\bd_T(S^*_T)\ge\bd(S^*)/2$, we have
\[ \f{w_T(E_T(S^*_T,V_T\sm S^*_T))}{\bd_T(S^*_T)} \le \f{w(E_{G_i}(S^*,V\sm S^*))}{\bd(S^*)/2}\le2\Psi_{G_i}^\bd(S^*)\le2\be \cd\psi^* .\]
Also, $\bd_T(S^*_T)\le2\bd(V)/3=2\bd_T(V_T)/3$.
Let $S'_T$ be the cut in $T$ that $\rooted$ outputs and let $S_T$ the corresponding cut in $G_i$ after the uncontraction in Step~\Dref{line:uncontract T}.
Applying \lemm{tree-alg} with $\psi_T=2\be \cd\psi^*$, the cut $S'_T$ satisfies
$\Psi_T^{\bd_T} (S'_T)\le 6\psi_T = 12\be \cd\psi^*$ and $\min\{\bd_T(S'_T),\bd_T(V_T\sm S'_T)\}\ge\bd_T(S^*_T)/3$. By construction, $\bd(S_T)=\bd(S'_T)\ge\bd_T(S^*_T)/3\ge\bd(S^*)/6 \ge \bd(S^*)/b$ and $\Psi_{G_i}^\bd(S_T)=\Psi_T^{\bd_T}(S'_T)\le12\be \cd\psi^* \le \psi$.

\im Suppose condition~(2) is true for the cut $S^*_H$. Since $w_H(E_{H}(S^*_H,K\sm S^*_H))\le w(E_{G_i}(S^*,V\sm S^*))$ and $\min \{ \bd_H(S^*_H) , \bd_H(K\sm S^*_H)\} \ge\bd(S^*)/3 $, we have $\Psi_H^{\bd_H}(S^*_H)\le3\Psi_{G_i}^{\bd}(S^*)\le3\be \cd\psi^*$. Since $H'$ is an $\al$-approximate spectral sparsifier of $H$, we have $\Psi_{H'}^{\bd_H}(S^*_H) \le \al \cd3\Psi_H^{\bd_H}(S^*_H)\le3\al \be \cd\psi^*$. By induction on the smaller recursive instance $\weighted(H',\bd_H, \psi/\al , 3\al \be \psi^*,b/3)$, the cut $S'_H$ computed is a $(3\al \be \psi^*,b/3)$-most-balanced $(\psi/\al ,\bd_H)$-sparse cut. Since $H'$ is an $\al$-approximate spectral sparsifier of $H$, we have $\Psi_H^{\bd_H}(S'_H) \le \al \cd\Psi_{H'}^{\bd_H}(S'_H)\le\al \cd \psi/\al =\psi$.
Let $S_H$ be the cut in $G_i$ corresponding to $S'_H$ after the uncontraction in Step~\Dref{line:uncontract H}.
By construction, $\Psi_{G_i}^\bd(S_H)=\Psi_H^{\bd_H}(S'_H) \le \psi$ and
$\bd(S_H)=\bd_H(S'_H)$. Since $S^*_H$ is a cut with $\Psi_{H'}^{\bd_H}(S^*_H)\le3\al \be \psi^*$, we have
\[ \bd(S_H)=\bd_H(S'_H)\ge\f{\min\{\bd_H(S^*_H),\bd_H(K\sm S^*_H)\}}{b/3} \ge \f{\bd(S^*)/3}{b/3}=\f{\bd(S^*)}b .\]
 \EE
In both cases, the computed cut is a  $(\psi/\psi^*,b)$-most-balanced $\psi $-sparse cut.
\EP

The lemma below will be useful in bounding the running time of the recursive algorithm.
\BL\leml{size-bound}
For any integer $t\ge1$ (as defined by the algorithm), the algorithm makes $t$ recursive calls $\weighted(H',\bd_H,\psi/\al ,\, 3\al \be \psi^*,\, b/3)$ on graphs $H'$ with $\tO(\f{m\log U}t)$ vertices and $\tO(\f{m \log^2U}t)$ edges, and runs in $\tO(tm)$ time outside these recursive calls.
\EL
\BP
By \thmm{madrythm}, computing the graphs $G_1,\lds,G_t$ takes $\tO(tm)$ time. By \lemm{tree-alg}, $\rooted$ runs in $O(m)$ time for each $G_t$, for a total of $O(tm)$ time. Since each graph $G_i$ is a $\tO(\f{m\log U}t)$-tree, by construction, each graph $H_{i}$ has at most $\tO(\f{m\log U}t)$ vertices. By \Cref{cor:sparsifier}, the sparsified graphs $H'_{i}$ have at most $\tO(\f{m\log U}t)\log m\log U\le\tO(\f{m\log^2U}t)$ edges.
\EP

Finally, we plug in our value $t = \left\lceil m^{1/r} (\log m)^{O(1)} \log^2 U \right\rceil$ that balances out the running time $\tO(tm)$ outside the recursive calls and the number $r$ of recursion levels.

\BT\thml{weighted-bal}
Fix parameters $\psi^*>0$ and $1\le r\le O(\log m)$, and let $\psi=12\be\cd(3\al^2\be)^r\cd\psi^*$.
There is a deterministic algorithm that, given a weighted graph $G$ with $m$ edges and capacity ratio $U$ and demands $\bd$, computes a $(12\be\cd(3\al^2\be)^r,6\cd3^r)$-most-balanced $\psi$-sparse cut in time $m^{1+1/r}\lp\log(mU)\rp^{O(1)}$. Note that $12\be\cd(3\al^2\be)^r = (\log m)^{O(r^3)}$.
\ET
\BP
Let $G$ be the original graph with $m=m_0$ edges.
Let $G'$ be the current input graph in a recursive call of $\weighted$, with $m'$ edges and capacity ratio $U'$. Set the parameters $t = \left\lceil m^{1/r} (\log m')^{O(1)} \log^2 U' \right\rceil$ from the algorithm and $\al =(\log m')^{O(r^2)}$ from \Cref{cor:sparsifier} and $\be =O(\log m'(\log\log m')^{O(1)})\le(\log m')^{O(1)}$ from \Cref{thm:madrythm}. By \lemm{size-bound}, the algorithm makes $t=m^{1/r} (\log m')^{O(1)}\log^2U'$ many recursive calls to graphs with at most  $\tO(\f{m'\log^2U'}t)\le m'/m^{1/r}$ edges, where $U'$ is the capacity ratio of the current graph, so there are $r$ levels of recursion. By \Cref{thm:madrythm}, the capacity ratio of the graph increases by an $O(m)$ factor in each recursive call, so we have $U'\le O(m)^r U$ for all recursive graphs, which means $t \le m^{1/r}(r\log m+\log U)^{O(1)}$. By \lemm{size-bound}, the running time $\tO(tm')$ outside the recursive calls for this graph is $m'm^{1/r}(r\log m+\log U)^{O(1)}$. For recursion level $1\le i\le r$, there are $m^{i/r}\lp r\log m+\log U\rp^{O(i)}$ many graphs at this recursion level, each with $m'\le m^{1-i/r}$, so the total time spent on graphs at this level, outside their own recursive calls, is at most
\[ m^{i/r}\lp r\log m+\log U\rp^{O(i)} \cd m^{1-i/r}m^{1/r}(r\log m+\log U)^{O(1)} = m^{1+1/r}(r\log m+\log U)^{O(i)} .\]
Summed over all $1\le i\le r$ and using $r\le O(\log m)$, the overall total running time becomes $m^{1+1/r}(\log(mU))^{O(r)}$.

We also need to verify that the conditions $\psi\ge12\be\cd\psi^*$ and $b\ge6$ of \lemm{analysis} are always satisfied throughout the recursive calls. Since each recursive call decreases the parameter $b$ by a factor of $3$, and $b=6\cd3^r$ initially, the value of $b$ is always at least $6$. Also, in  each recursive call, the ratio $\psi/\psi^*$ decreases by a factor $3\al^2\be$, so for the initial value $\psi=12\be\cd(3\al^2\be)^r\cd\psi^* $ in the theorem statement, we always have $\psi/\psi^*\ge12\be$.
\EP

\subsection{Completing the Proof of \thmm{WBCut} and \Cref{cor:exp-decomp-sec8}} \label{sec:completing}

The proofs in this section follow the template from \cite{NanongkaiS17} but generalize it to work in weighted graphs and general demand.
In order to prove \thmm{WBCut}, we first present the lemma below.
Roughly, it guarantee the following. Given a set $V'$ where $G[V']$ is ``close'' to being an expander in the sense that any sparse cut $(A,B)$ in $V'$ must be unbalanced: $\min\set{\bd(A),\bd(B)}\leq z$, then the algorithm returns a large subset $Y \subseteq V'$ such that $Y$ is ``closer'' to being an expander. That is, any sparse cut $(A',B')$ in $Y$ must be even more unbalanced: $\min\set{\bd(A'),\bd(B')}\leq z' \ll z$.

\begin{lem}\leml{z}
Let $G=(V,E)$ be a weighted graph with edge weights in $[1,U]$, and demands $\bd(v)\in\{0\}\cup[1,U]$ for all $v\in V$ that are not all zero.
        There is a universal constant $c_1>0$ and a deterministic algorithm, that, given a vertex subset $V'\subseteq V$ with $\bd(V')\ge\bd(V)/2$,
        and parameters $r\ge1$, $0<\psi<1$, $0<z'<z$, such that for every partition $(A,B)$ of $V'$ with $w(E_{G}(A,B))\leq \psi\cdot  \min\set{\bd(A),\bd(B)}$, $\min\set{\bd(A),\bd(B)}\leq z$ holds,
        computes a partition $(X,Y)$ of $V'$, where  $\bd(X)\leq \bd(Y)$ (where possibly $X=\emptyset$),  $w(E_{G}(X,Y))\leq \psi \cdot\bd(X)$, and one of the following holds:

        \begin{enumerate}
                \item either $\bd(X),\bd(Y)\geq \bd(V')/3$ (note that this can only happen if $z\geq \bd(V')/3$); or
                \item for every partition $(A',B')$ of the set $Y$ of vertices with \[ w(E_{G}(A',B'))\leq \frac{\psi}{(\log(mU))^{c_1 r^3}}\cdot \min\set{\bd(A'),\bd(B')},\] $\min\set{\bd(A'),\bd(B')}\leq z'$ must hold (if $z'<1$, then graph $G[Y]$ is guaranteed to have $\bd$-sparsity at least $\psi / (\log(mU))^{c_1 r^3}$). \label{item:2}
        \end{enumerate}
The running time of the algorithm is $O\left (\frac{z}{z'}\cdot m^{1+1/r}\lp\log(mU)\rp^{O(1)} \right )$.
\end{lem}
\BP

Our algorithm is iterative. At the beginning of iteration $i$, we are given a subgraph $G_i\subseteq  G$, such that $\bd(V(G_i))\geq 2\bd(V')/3$; at the beginning of the first iteration, we set $G_1= G[V']$.
At the end of iteration $i$, we either terminate the algorithm with the desired solution, or we compute a subset $S_i\subseteq V(G_i)$ of vertices, such that $\bd(S_i)\leq \bd(V(G_i))/2$, and $w(E_{G_i}(S_i, V(G_i)\setminus S_i))\leq \psi\cdot \bd(S_i)/2$. We then delete the vertices of $S_i$ from $G_i$, in order to obtain the graph $G_{i+1}$, that serves as the input to the next iteration. The algorithm terminates once the current graph $G_i$ satisfies $\bd(V(G_i)) < 2\bd(V')/3$ (unless it terminates with the desired output beforehand).

We now describe the execution of the $i$th iteration. We assume that the sets $S_1,\ldots,S_{i-1}$ of vertices are already computed, and that $\sum_{i'=1}^{i-1}\bd(S_{i'})\leq \bd(V')/3$. Recall that $G_i$ is the sub-graph of $ G[V']$ that is obtained by deleting the vertices of $S_1,\ldots,S_{i-1}$ from it. Recall also that we are guaranteed that $\bd(V(G_i))\geq 2\bd(V')/3\geq \bd(V)/3$. We apply \thmm{weighted-bal} to graph $G_i$ with parameters $\psi^*=(\psi/2) / (\log(mU))^{c_1 r^3}$ and $r$, and let $X$ be the returned set, which is  a  $(\psi^*,6\cd3^r)$-most-balanced $((\log(mU))^{c_1 r^3} \cd \psi^*,\bd)$-sparse cut satisfying $\bd(X)\le\bd(V)/2$.

We set parameter $z^*=z'/(6\cd3^r)$. If $\bd(X)\le z^*$, then we terminate the algorithm, and return the partition $(X,Y)$ of $V'$ where $X=\bigcup_{i'=1}^iS_{i'}$, and $Y=V'\setminus X$. This satisfies the second condition of \lemm{z}, since by the most-balanced sparse cut definition, every partition $(A',B')$ of the set $Y$ of vertices with $w(E_{\hG}(A',B'))\leq \frac{\psi}{(\log(mU))^{c_1 r^3}}\cdot \min\set{\bd(A'),\bd(B')}$ must satisfy $\min\{\bd(A'),\bd(B')\}\le6\cd3^r\cd\bd(X) < 6\cd3^r\cd z^*=z'$.

Otherwise, $\bd(X)> z^*$. In this case, we set $S_i=X$ and continue the algorithm. If $\sum_{i'=1}^i\bd(S_{i'})\leq \bd(V')/3$ continues to hold, then we let $G_{i+1}=G_i\setminus S_i$, and continue to the next iteration. Otherwise, we terminate the algorithm, and return the partition $(X,Y)$ of $V'$ where $X=\bigcup_{i'=1}^iS_{i'}$, and $Y=V'\setminus X$. Recall that we are guaranteed that $\bd(X)\geq \bd(V')/3$.

To show that  $w(E_{\hG}(X,Y))\leq \psi \cdot\bd(X)$, note that every cut $S_i$ satisfies $w(E_{G_{i}}(S_i,V(G_i)\sm S_i))\le(\psi/2)\bd(S_i)$, so $w(E_G(X,Y)) \le \sum_{i'=1}^iw(E_{G_i}(S_i,V(G_i)\sm S_i)) \le (\psi/2)\sum_{i'=1}^i\bd(S_i)=(\psi/2)\bd(X)$, which is at most $\psi\min\{\bd(X),\bd(V\sm X)\}$ since  $\bd(X)\le2\bd(V)/3$.

The bound on the running time of the algorithm proceeds similarly. Observe that we are guaranteed that for all $i$, $\bd(S_i)\geq z^*$. Notice however that throughout the algorithm, if we set $A=\bigcup_{i'=1}^iS_{i'}$ and $B=V'\setminus A$, then
$\bd(A)<\bd(B)$ holds, and $w(E_{ G}(A,B))\leq \psi\cdot  \bd(A)$. Therefore, from the condition of the lemma, $\bd(A)\leq z$ must hold. Overall, the number of iterations in the algorithm is bounded by $z/z^*=6\cd3^r\cdot z/z'$, and, since every iteration takes time $m^{1+1/r}\lp\log(mU)\rp^{O(1)} $, total running time of the algorithm is bounded by $\frac z {z'}\cdot m^{1+1/r}\cdot (\log (mU))^{O(1)}$.
\EP

We are now ready to complete the proof of \thmm{WBCut}, which is almost identical to the proof of Theorem~7.5 of~\cite{ChuzhoyGLNPS20}. For completeness, we include the proof below. %

\begin{proof}[Proof (\thmm{WBCut})]
We first show that we can safely assume that $\bd(V)\ge 2\cd4^r$. Otherwise, consider the following expression in \ref{item:2} of \Cref{lem:z} and its upper bound:
\[ \frac{\psi}{(\log(mU))^{c_1 r^3}}\cdot \min\set{\bd(A'),\bd(B')} \le \frac{1}{(\log(mU))^{c_1 r^3}}\cdot 2\cd 4^r < 1, \]
which holds for large enough $c_1>0$. Since $G$ is connected and all edges have weight at least $1$, the condition in \ref{item:2} only applies with $A'=\emptyset$ or $B'=\emptyset$. Therefore, the algorithm can trivially return $X=\emptyset$ and $Y=V$ and satisfy \ref{item:2}.

For the rest of the proof, assume that $\bd(v)\ge2\cd4^r$. Our algorithm will consist of at most $r$ iterations and uses the following parameters. First, we set $z_1=\bd(V)/2$, and for $1<i\leq r$, we set $z_i=z_{i-1}/(\bd(V)/2)^{1/r} \le z_{i-1}/4$; in particular, $z_r=1$ holds. We also define parameters $\psi_1,\ldots,\psi_{r}$, by letting $\psi_r=\psi$, and, for all $1\leq i<r$, setting $\psi_{i}=8 \cd (\log(mU))^{c_1 r^3}\cdot \psi_{i+1}$, where $c_1$ is the constant from \lemm{z}. Notice that $\psi_1\leq \psi \cd  (\log m)^{O(r^4)}$. 

In the first iteration, we apply \ref{lem:z} to the set $V'=V$ of vertices, with the parameters $\psi=\psi_1$, $z=z_1$, and $z'=z_2$.
Clearly, for every partition $(A,B)$ of $V'$ with $w_G(E_{ G}(A,B))\leq \psi_1\cdot \min\set{\bd(A),\bd(B)}$, it holds that $\min\set{\bd(A),\bd(B)}\leq z_1=\bd(V)/2$.
 If the outcome of the algorithm from  \ref{lem:z} is a partition $(X,Y)$ of $V$ satisfying $\bd(X),\bd(Y)\geq \bd(V)/3$ and $w_G(E_{ G }(X,Y))\leq \psi_1\cdot \min\set{\bd(X),\bd(Y)}\leq \psi\cdot (\log m)^{O(r^4)}\min\set{\bd(X),\bd(Y)}$, then we return the cut $(X,Y)$ and terminate the algorithm. 

We assume from now on that the algorithm from \ref{lem:z} returned a partition $(X,Y)$ of $V$, where $\bd(X)\leq \bd(Y)$ (where possibly $X=\emptyset$), $\bd(X)\le\bd(V)/3$, $w_G(E_{ G }(X,Y))\leq \psi_1 \cdot\bd(X)$, and the following guarantee holds: For every partition $(A',B')$ of the set $Y$ of vertices with $w_G(E_{ G }(A',B'))\leq 8\psi_2\cdot \min\set{\bd(A'),\bd(B')}$, it holds that $\min\set{\bd(A'),\bd(B')}\leq z_2$. We set $S_1=X$, and we let $ G_2= G \setminus S_1$.

The remainder of the algorithm consists of $r-1$ iterations $i=2,3,\lds,r$. The input to iteration $i$ is a subgraph $ G_i\subseteq  G$ with $\bd(V(G_i))\le\bd(V)/2$, such that for every cut $(A',B')$ of $ G_i$ with $w_G(E_{ G }(A',B'))\leq \psi_i\cdot \min\set{\bd(A'),\bd(B')}$, it holds that $\min\set{\bd(A'),\bd(B')}\leq z_i$. (Observe that, as established above, this condition holds for graph $ G_2$). 
The output is a subset $S_{i}\subseteq V( G_i)$ of vertices, such that $\bd(S_i)\le\bd(V(G_i))/2$ and $w_G(E_{ G_i}(S_i,V( G_i)\setminus S_i))\leq \psi_i \cdot \bd(S_i)$, and, if we set $ G_{i+1}= G_i\setminus S_{i}$, then we are guaranteed that for every cut $(A'',B'')$ of $ G_{i+1}$ with $w_G(E_{ G }(A'',B''))\leq 8\psi_{i+1}\cdot \min\set{\bd(A''),\bd(B'')}$, it holds that $\min\set{\bd(A''),\bd(B'')}\leq z_{i+1}$.
In particular, if $w_G(E_{ G }(A'',B''))\leq \psi_{i+1}\cdot \min\set{\bd(A''),\bd(B'')}$, then $\min\set{\bd(A''),\bd(B'')}\leq z_{i+1}$ holds.
 In order to execute the $i$th iteration, we simply apply \ref{lem:z} to the set $V'=V( G_i)$ of vertices, with parameters $\psi=\psi_i$, $z=z_i$ and $z'=z_{i+1}$. As we show later, we will ensure that $\bd(V(G_i))\geq \bd(V)/2$. Since, for $i>1$, $z_i\le\bd(V)/8<\bd(V)/6\leq \bd(V(G_i))/3$, the outcome of the lemma must be a partition $(X,Y)$ of $V'$, where $\bd(X)\leq \bd(Y)$ (where possibly $X=\emptyset$),  $w_G(E_{ G }(X,Y))\leq \psi_i \cdot\bd(X)$, and we are guaranteed that, for every partition $(A'',B'')$ of the set $Y$ of vertices with $w_G(E_{ G }(A'',B''))\leq 8\psi_{i+1}\cdot \min\set{\bd(A''),\bd(B'')}$, it holds that $\min\set{\bd(A'),\bd(B')}\leq z_{i+1}$. Therefore, we can simply set $S_i=X$, $ G_{i+1}= G_i\setminus S_i$, and continue to the next iteration, provided that $\bd(V( G_{i+1}))\geq \bd(V)/2$ holds.

 We next show that this indeed must be the case. Recall that for all $2\leq i'\leq i$, we guarantee that $\bd(S_{i'})\le z_{i'}\le\bd(V)/(2\cd4^{i'-1})$.
 Therefore, if we denote by $Z=\bigcup_{i'=2}^iS_{i'}$ and $Z'=V( G_2)\setminus Z$, then $\bd(Z)\le\bd(V)/2 \cd \sum_{i'=2}^i1/4^{i'-1}\le \bd(V)/6$, so 
\[ \bd(V(G_{i+1}))=\bd(Z')=\bd(V(G_2))-\bd(Z)\ge2 \bd(V)/3-\bd(V)/6=\bd(V)/2 .\] 
as promised.

 We continue the algorithm until we reach the last iteration, where $z_r=1$ holds. Apply  \ref{lem:z} to the final graph $ G_r$ with $z'=1/2$ to obtain $S_r\s V(G_r)$. Since $z'<1$, the discussion in \ref{item:2} implies that graph $ G_r\setminus S_r $ has $\bd$-sparsity at least $\psi$ (recall that $\psi_r=\psi$). We define our final partition as $Y=V(G_r)\sm S_r$ and $X=V\sm Y=\bigcup_{i=1}^rS_i$. By the same reasoning as before, we are guaranteed that $\bd(Y)\ge\bd(V)/2\ge\bd(X)$. Finally,
\[ w_G(E_G(X,Y))\le \sum_{i=1}^rw_G(E_G(S_i,V(G_i)\sm S_i))\le\sum_{i=1}^r\psi_i\cd\bd(S_i)\le\psi \cd  (\log m)^{O(r^4)}\cd\bd(X), \]
which concludes the proof of \thmm{WBCut}.
\end{proof}

Finally, we prove \Cref{cor:exp-decomp-sec8}, which is almost identical to the proof of Corollary~8.5 of~\cite{ChuzhoyGLNPS20}.  \begin{proof}[Proof (\Cref{cor:exp-decomp-sec8})]
        We maintain a collection $\hset$ of disjoint sub-graphs of $G$ that we call \emph{clusters}, which is partitioned into two subsets, set $\hset^A$ of \emph{active clusters}, and set $\hset^I$ of \emph{inactive clusters}. We ensure that each inactive cluster $H\in \hset^I$ has $\bd|_{V(H)}$-sparsity at least $\psi$. We also maintain a set $E'$ of ``deleted'' edges, that are not contained in any cluster in $\hset$. At the beginning of the algorithm, we let $\hset=\hset^A=\set{G}$, $\hset^I=\emptyset$, and $E'=\emptyset$. The algorithm proceeds as long as $\hset^A\neq \emptyset$, and consists of iterations.
        For convenience, we denote $\alpha=(\log m)^{O(r^4)}$ the approximation factor achieved by the algorithm from \thmm{WBCut}, and we set $\psi=\epsilon/(c\alpha\cdot \log (mU))$, for some large enough constant $c$, so that
$\psi=\Omega\left(\epsilon/ \big( \log^{O(r^4)}m \log U \big)\right)$ holds.

        In every iteration, we apply the algorithm from \Cref{thm:WBCut} to every graph $H\in \hset^A$, with the same parameters $\alpha$, $r$, and $\psi$. Consider the partition $(A,B)$ of $V(H)$ that the algorithm computes, with $w(E_H(A,B))\leq \alpha \psi\cdot \bd(V(H))\leq \frac{\epsilon\cdot \bd(V(H))}{c\log (mU)}$. We add the edges of $E_H(A,B)$ to set $E'$. If $\bd(A),\bd(B)\ge \bd(V(H))/3$, then we replace $H$ with $H[A]$ and $H[B]$ in $\hset$ and in $\hset^A$. Otherwise, we are guaranteed that
        $\bd(A)\ge\bd(V(H))/2$ and $\Psi^{\bd|_A}(H[A])\ge\psi$. Then we remove $H$ from $\hset$ and $\hset^A$, add $H[A]$ to $\hset$ and $\hset^I$, and add $H[B]$ to $\hset$ and $\hset^A$.

        When the algorithm terminates, $\hset^A=\emptyset$, and so every graph $H\in \hset$ has $\bd|_{V(H)}$-sparsity at least $\psi$. Notice that in every iteration, the maximum value of $\bd(V(H))$ of a graph $H\in\hset^A$ must decrease by a constant factor. Therefore, the number of iterations is bounded by $O(\log(mU))$. It is easy to verify that the total weight of edges added to set $E'$ in every iteration is at most $\frac{\epsilon\cdot \bd(V)}{c\log (mU)}$. Therefore, by letting $c$ be a large enough constant, we can ensure that $w(E')\leq \epsilon\bd(V)$. The output of the algorithm is the partition $\pset=\set{V(H)\mid H\in \hset}$ of $V$. From the above discussion, we obtain a valid $(\epsilon, \psi)$-expander decomposition, for $\psi=\Omega\left(\epsilon/ \big( \log^{O(r^4)}m \log U \big)\right)$.

        It remains to analyze the running time of the algorithm.  The running time of a single iteration is bounded by $m\cd(mU)^{O(1/r)} $. Since the total number of iterations is bounded by $O(\log (mU))$, we get that the total running time of the algorithm is $m\cd(mU)^{O(1/r)}\log(mU)$.
\end{proof}

\section*{Acknowledgements} 
We thank Julia Chuzhoy and Richard Peng for helping improving the presentation of this note and helpful comments.

\bibliographystyle{alpha}
\bibliography{bibliography}

\end{document}